\documentstyle[11pt]{article}

\begin{document}

\begin{center}
{\Large\bf Theory of sound propagation in superfluid-filled porous media }

\vspace{0.5in}

{ T.Buishvili }

{Dept.\ of Physics, Tbilisi State University, Chavchavadze
ave.\ 3,\\ Tbilisi 380028, Georgia,}
\vspace{0.2in}

{Sh.Kekutia, O.Tkeshelashvili}

{Institute of Cybernetics, Sandro Euli str.\ 5,
 Tbilisi 380086, Georgia,}
\vspace{0.2in}

{L.Tkeshelashvili}

{Institut f\"ur Theorie der
Kondensierten Materie,
University of Karlsruhe,
P.O. Box 6980, 76128 Karlsruhe, Germany;
\\ Institute of
Physics, Tamarashvili str.\ 6, Tbilisi 380077, Georgia,
\\ E-mail: lasha@tkm.physik.uni-karlsruhe.de}
\date{\today}

\end{center}
\vspace{0.5in}
\begin{abstract}

The theory of sound propagation in macroscopically  isotropic and homogeneous 
porous media saturated with 
superfluid $^4$He  has been developed
neglecting all damping processes.  The case when the normal fluid
component is locked inside a porous medium by viscous forces is
investigated in detail. It is shown that in this case one shear wave  and
two longitudinal, fast and slow, waves exist. 
 Fast wave as well as slow wave is accompanied with  temperature oscillations. The velocities of these  waves are obtained.

\end{abstract}

\vspace{5mm}


KEY words: porous materials; hydrodynamical waves;
\newpage

\section{Introduction}

The investigation  of acoustic propagation in a fluid-saturated porous media
provides opportunities to study  disordered pore structures, as well as the
properties of the fluids confined in the pores. In [1] Biot developed a simple theory of propagation of elastic
waves in statistically homogeneous and isotropic porous media saturated
with a (one-component) compressional fluid 
  assuming that pore
sizes are smaller  compared to the wavelength i.e. the long wavelength limit
is considered. Biot's theory predicts  one shear wave and two longitudinal,
fast and slow compressional waves. The fast  compressional wave corresponds
to solid and fluid moving in phase, and the slow one
corresponds to solid and fluid moving in opposite phase. 
The Biot's theory was experimentally confirmed by the
observation of the fast and slow
compressional wave in glass bead samples filled with water[2].

Because of the unique properties
of liquid helium [3], it is of fundamental importance
 to investigate the sound
propagation in superfluid-filled porous media. Below the $\lambda$-transition temperature $T_\lambda=2.17 K$, liquid $^{4}$He
becomes a superfluid and behaves as a ``mixture'' of a normal(viscous) component
with mass density $\rho_n$ and a superfluid (zero-viscosity) component with
mass density $\rho_s$; 
it is clear that $\rho_n+\rho_s=\rho_h$ where $\rho_h$
is the mass density of superfluid $^{4}$He(so-called He II). It must be pointed out that the superfluid component and the normal component can not be physically separated [3].   The
normal (superfluid) component fraction increases(decreases) monotonously
from $\rho_n=0$($\rho_s=\rho_h$) at $T=0$ to $\rho_n=\rho_h$($\rho_s=0$)
at and above $T_\lambda$. It is important that $\rho_n$ and $\rho_s$ can move
independently, and therefore in bulk He II the two mode of
sound propagation are possible. In first sound the two fluid components
oscillate in phase; and in second sound $\rho_s$ and $\rho_n$ oscillate
in opposite phase. One of the most interesting property of He II is fourth
sound mode. In fourth sound  only the superfluid component moves, when the
normal fluid component is locked inside a rigid porous medium by viscous
forces, so that its velocity is identically zero.

Since a superfluid-filled porous medium is really a three-component
system, Biot's theory is not applicable  directly to this system. But, at low
temperatures ($T<1.1 K$), when the normal fluid component fraction may be
neglected, Johnson has shown that fourth sound mode is exactly a  Biot's slow
wave [4]. It should be  noted that at low
temperatures, when the mean free path of
phonons and rotons becomes larger than the pore size the
hydrodynamical description breaks down, so the comparison to Biot's
model should be made at higher temperatures [5].

We have to  point out that a  number of authors extended Biot's theory in the case of two or more fluids (for review see [6],[7]). Since the fluids usually considered are not superfluids, the  authors neglected some effects, which are actually important in the case of superfluids. Indeed, it is well known  that in the case of ordinary fluids the fountain(and/or mechano-caloric) effect is  negligible, however, in  the case of a superfluid these effects are crucial. In particular, a temperature gradient produces relative motion between the two fluids in superfluids [3]. Therefore  we have to distinguish between pressures with a ``mechanical origin'' and those with a ``thermal origin'', the later can only be found in superfluids [3]  whereas   the former can appear in any fluid. Moreover, it is important that the superfluid component and the normal component can not be physically separated. Actually He II is the fluid with additional degrees of freedom [3], but not a real  mixture of  two fluids.

In this paper  the  generalization of Biot's theory is given
in the case of superfluid-saturated
porous media, when all damping processes are neglected. The organization of this paper is as follows. In Section II we apply Biot formalism to  bulk He II and derive some equations, which are used in Section IV. Section III and Section IV give  a derivation of the equations which describe the propagation of longitudinal and shear   waves in a superfluid-saturated, homogeneous and isotropic porous media. This is the  main result of this paper. In Section V the formalism is applied to the most important particular  case when the normal fluid   component is locked inside a porous media by viscous  forces. Finally, in Section VI we discuss some limits of our equations and  summarize  results.       

\section{The case of bulk He II}

In this section  we apply formalism developed by Biot [1],[8] to the wave processes in bulk He II. The wave processes in bulk He II is well understood [3], but Biot's formalism  has the advantage that it can be generalized easily for three-component systems. Besides, in this section  we derive some equations which are used later.   

The basic idea is to describe the wave processes in He II in terms of ``elastic deformations'' of the superfluid and the  normal components of He II. In this formalism the pressure with a "thermal origin" have to be considered as an ``elastic'' restore force as well as the pressure with a "mechanical origin". Here we present the key points of calculations, for more discussion of details of the formalism the  reader is referred to [1],[8].

Let us  define the displacement of the superfluid and the normal components as $\vec u_s$ and $\vec u_n$ respectively. Then the displacement $\vec u$ of He II is [3]:

\begin{equation}
\vec u=\frac{\rho_s}{\rho_h}\vec u_s+\frac{\rho_n}{\rho_n}\vec u_n.
\end{equation}

Following [1] we choose the Lagrangian coordinates of the system as the six displacement components $ u_{sx}, u_{sy}, u_{sz}, u_{nx}, u_{ny}, u_{nz}$ of He II. Note that in contrast with ref.[1] no averaging  procedure is  required here.
Then the kinetic energy $T_h$ of the unit volume of He II is presented as:

\begin{equation}
2T_h=\sum_{x,y,z}\left(\rho_n\left[\frac{\partial u_{ni}}{\partial t}\right]^2+\rho_s\left[\frac{\partial u_{si}}{\partial t}\right]^2\right);
\end{equation}
here we take into account that the superfluid component and the normal component move independently. So the  mass coupling parameter between the  normal component and the  superfluid component equals zero [1].

Let us assume that the system is in equilibrium being at rest. Therefore, any displacement is a deviation from a state of minimum potential energy. Then  the potential energy $W_h$ per unit volume of He II is [1],[9]:
\begin{equation}
2W_h=\sigma_n^h\epsilon_n+\sigma_s^h\epsilon_s;
\end{equation} 
here we take into account that the shear modulus of He II is zero. We formally separate the stress of the normal component  $\sigma_n^h$ and  the stress of the superfluid component  $\sigma_s^h$ (see Section III as well); 
they describe the pressure with a "thermal origin" as well as the pressure with a "mechanical origin''. $\epsilon_s=div\ \vec{u}_s$ and $\epsilon_n=div\ \vec{u}_n$ are the strain in the normal and the superfluid components of He II respectively. From (1) the  strain in the fluid $\epsilon$ is given by:
\begin{equation}
\epsilon=\frac{\rho_s}{\rho_h}\epsilon_s+
\frac{\rho_n}{\rho_h}\epsilon_n.
\end{equation}

When there is no (nonuniform) relative motion between the superfluid and the normal components then  $\epsilon_s=\epsilon_n$ and no temperature gradients exist (see the text after equation (16)). In this case the dynamics of the system is described by the  theory of elasticity. 

But when $\epsilon_s\not=\epsilon_n$  temperature gradients exist and the fountain and/or mechano-caloric effect takes place (see the text after equation (16)). We show in this section that in this case the formalism of theory of elasticity is applicable as well.

In  the first approximation the stress components are linear functions of the strain components, so one gets [1]:
\begin{eqnarray}
\sigma_s^h&=&R_s^h\epsilon_s+R_{sn}^h\epsilon_n;\nonumber\\
\sigma_n^h&=&R_n^h\epsilon_n+R_{sn}^h\epsilon_s;
\end{eqnarray} 
$R_s^h$ and $R_n^h$ describe "elastic properties'' of the superfluid and the normal components, and $R_{sn}^h$ describes the ``elastic interaction'' between them [1]. 

To make results more useful  let us express $R_s^h$, $R_n^h$ and $R_{sn}^h$ in terms of directly measurable coefficients [8]. 

Let us now  consider a volume of He II in a vessel. First we consider the case,  when none of the fluid components  is allowed to escape from vessel. We shall refer to  this case as ``unjacketed compressibility test''(see Section III). If a pressure $p$ is applied to He II (see Fig. 1), no relative  motion  between the superfluid component and the normal component takes place, i.e.:
\begin{equation}
\epsilon_s=\epsilon_n=\epsilon.
\end{equation}

So, in this case temperature  difference does not exist. Fluid pressure has only ``mechanical origin'' and  we have [8]:
\begin{eqnarray}
\sigma_s^h&=&-\frac{\rho_s}{\rho_h}p;\nonumber\\
\sigma_n^h&=&-\frac{\rho_n}{\rho_h}p;
\end{eqnarray}
so, from (5) and (7) we get:
\begin{eqnarray}
-\frac{\rho_s}{\rho_h}p&=&R_s^h\epsilon_s+R_{sn}^h\epsilon_n;\nonumber\\
-\frac{\rho_n}{\rho_h}p&=&R_n^h\epsilon_n+R_{sn}^h\epsilon_s;
\end{eqnarray}

Besides, in this case  we have:
\begin{equation}
\frac{1}{K_f}=-\frac{\epsilon}{p};
\end{equation}
where $K_f$ is the adiabatic  bulk modulus of He II.

Using (9) from (8) we get (recall that $\epsilon=\epsilon_n$, in this case):
\begin{eqnarray}
R_s^h&=&\frac{\rho_s}{\rho_h}K_f-R_{sn}^h.\nonumber\\
R_n^h&=&\frac{\rho_n}{\rho_h}K_f-R_{sn}^h.
\end{eqnarray}

Now we consider another case, when the superfluid component is allowed to escape   through a superleak, but not the normal fluid component (see Fig. 2). We shall refer to  this case as ``second jacketed compressibility test'' (see Section III).  If a pressure $p$ is     applied  to He II, the superfluid component passes through the superleak into     the reservoir of He II. An applied external pressure squeeze out some of the superfluid component, this means that  temperature increases in a vessel.  So, a  temperature difference between the vessel and the reservoir of He II  exists. As it  is well known,  temperature difference between   the two volumes of He II is accompanied by a pressure difference between these volumes (fountain or/and  mechano-caloric effect)[3]. Since  the  superfluid component is  allowed to  pass through the  superleak,   its partial pressure is zero.  But the normal fluid component is identified  with the thermal excitations of He II (gas of phonons and rotons) and its partial pressure  is not zero. Thus, if we neglect the thermal expansion coefficient of He II (which is anomalously small), we can conclude that fluid pressure in this case  has   only ``thermal origin''. It is clear that the normal  fluid component  pressure equals  to the  external pressure (let us define it in this case by $p'$), and we have:
\begin{eqnarray}
\sigma_s^h&=&0;\nonumber\\
\sigma_n^h&=&-p';
\end{eqnarray}
or using (5) and (11):
\begin{eqnarray}
0&=&R_s^h\epsilon_s+R_{sn}^h\epsilon_n;\nonumber\\
-p'&=&R_n^h\epsilon_n+R_{sn}^h\epsilon_s;
\end{eqnarray}

Let us define the bulk modulus of the normal fluid component $K_n$ as:
\begin{equation} 
\frac{1}{K_n}=-\frac{\epsilon_n}{p'};
\end{equation}
Since the superfluid component and the normal component can not be physically separated, the definition of the bulk modulus of the  normal fluid component is formal. But let us point out that $K_n$ does not enter into final equations (18), (19); so this problem is irrelevant in this case. 

From (13) and (12) we have:

\begin{equation}
(R_{sn}^h)^2+R_s^h K_n\left(1-\frac{R_n^h}{K_n}\right)=0.
\end{equation}

Besides, in this case the energy conservation law gives:
\begin{equation}
-\frac{\rho_s}{\rho_h}(\epsilon_n-\epsilon_s)S\rho_hT=\rho_hC\Delta T;
\end{equation}
where $(\rho_s/\rho_h)(\epsilon_n-\epsilon_s)$ represents the fluid volume entering through the superleak [8]. $C$ and $S$ are the specific heat and the entropy per unit mass of He II, respectively. $\Delta T$ is the difference between the temperature of the fluid inside a vessel and the temperature of the reservoir of He II caused by external pressure [3]:
\begin{equation}
\Delta T=\frac{p'}{\rho_hS}.
\end{equation}  
One can see from (15) that $\Delta T=0$  when $\epsilon_s-\epsilon_n=0$, so pressure has only a ``mechanical origin''. But when $\epsilon_s\not=\epsilon_n$ from (15) one gets that $\Delta T\not=0$, therefore temperature gradients exist. 

From (15)  using the first equation in (12) we can eliminate $\epsilon_s$ and $\epsilon_n$, and we get the following  expression for $K_n$:
\begin{equation}
\frac{1}{K_n}=\frac{1}{K_f}+\frac{1}{\rho_h}\frac{C}{S^2T};
\end{equation}
here we have used (10), (14), (16) and definition (13) as well.

From (10) and (14)  using (17) we directly get:   
\begin{eqnarray}
R_s^h&=&K_f\ \frac{\rho_s^2}{\rho_h^2}\ +\frac{\rho_hS^2T}{C}\frac{\rho_s^2}{\rho_h^2};\nonumber\\
R_n^h&=&K_f\ \frac{\rho_n^2}{\rho_h^2}\ +\frac{\rho_hS^2T}{C}\frac{\rho_s^2}{\rho_h^2};\nonumber\\ 
R_{sn}^h&=&K_f\frac{\rho_s\rho_n}{\rho_h^2}-\frac{\rho_hS^2T}{C}\frac{\rho_s^2}{\rho_h^2}.
\end{eqnarray}

Equation (18) gives  $R_s^h$, $R_n^h$ and $R_{sn}^h$ in terms of directly measurable coefficients. These relations  are used in Section IV.

Now, to demonstrate  the correctness of the  expressions obtained above let us calculate the velocity of sound propagation in He II. 

Using (2), (3) and (5) from Lagrange's equations one derives the equations for the wave propagation in He II [1]: 
\begin{eqnarray}
\rho_s\frac{\partial^2}{\partial t^2}\epsilon_s&=&\vec\nabla^2(R_s^h\epsilon_s+R_{sn}^h\epsilon_n); \nonumber\\
\rho_n\frac{\partial^2}{\partial t^2}\epsilon_n&=&\vec\nabla^2(R_{sn}^h\epsilon_s+R_n^h\epsilon_n);
\end{eqnarray}
Solutions of (19) with $\vec k$ wave vector have the form:
\begin{eqnarray}
\epsilon_s&=&C_s \exp[i\vec k(\vec r+\vec Vt)];\nonumber\\
\epsilon_n&=&C_n \exp[i\vec k(\vec r+\vec Vt)];
\end{eqnarray}
$C_s$ and $C_n$ are real constants, and $\vec V$ is the velocity of the wave. 

Substituting (20) into (19)  we find that there are two modes of longitudinal waves the fast and the slow waves  with velocities  $V_1$ and $V_2$ respectively. Using (18) we get the expressions for $V_1$ and $V_2$:
\begin{eqnarray}
V_1^2&=&\frac{K_f}{\rho_h};\nonumber\\
V_2^2&=&\frac{\rho_s}{\rho_n}\frac{S^2T}{C};
\end{eqnarray}
 The fast wave corresponds to the superfluid component and the normal component moving in phase; from (19) one can see that  in the case of  fast wave $C_s\ =\ C_n$ [1]. This means that  $\epsilon_s=\epsilon_n$, so no fountain(and/or mechano-caloric) effect takes place and only the pressure with a ``mechanical origin'' appear. This is a ordinary wave in an ordinary fluid  (see also Section V). 

 The slow wave corresponds to the superfluid and the normal components moving in opposite phase. In the case of slow wave  (19) gives that  $\rho_s C_s+\rho_n C_n=0$ [1]. Thus $\epsilon=0$ and  pressure with a ``mechanical origin'' does not appear; but  $\epsilon_s\not=\epsilon_n$ , thus temperature oscillations exist. This is a temperature wave which appear only in superfluids (see also Section V). 

These are  well known results of Landau's theory of sound propagation in He II [3]. Therefore, we have  shown that the  formalism developed by Biot [1] may be applied to the wave processes in He II. Besides, we conclude that bulk He II may be regarded as a fluid-saturated porous  media with porosity $\rho_s/\rho_h$. The normal fluid component plays the role of the solid part with $K_n$ bulk modulus and zero shear modulus, and the superfluid component plays the role of the fluid. Equation  (21) shows that second sound is exactly slow wave mode.       

\section{Derivation of general equations}

In this section  we will derive the general equations  which govern the propagation of longitudinal and shear waves in a superfluid-saturated, homogeneous and isotropic porous media.

Let us consider a unit cube of superfluid-filled, macroscopically homogeneous
and isotropic porous solid as an element. The element is assumed to be large
as compared to pore sizes, but small compared to a wavelength. In this case,
the element is described by the average displacement of the normal fluid
component $\vec{u}_n$, the superfluid component $\vec{u}_s$, and the solid
$\vec{u}$ parts.

Following [1], let us introduce the Lagrangian viewpoint
and the concept of
 generalized coordinates. The Lagrangian coordinates are chosen as the nine
average displacement components of both 
the solid and the fluid i.e. $u_x, u_y, u_z,
 u_{sx}, u_{sy}, u_{sz}, u_{nx}, u_{ny}, u_{nz}.$

The kinetic energy $T$ of the system per unit volume is [1]:
\begin{eqnarray}
2T&=&\sum_{i=x,y,z}\left(\rho_{11}\left[\frac{\partial u_i}{\partial t}\right]^2
+\rho_{22}^s\left[\frac{\partial u_{si}}{\partial t}\right]^2+
\rho_{22}^n\left[\frac{\partial u_{ni}}{\partial t}\right]^2+\right.\nonumber\\
& &\left.2\rho_{12}^s\left[\frac{\partial u_{i}}{\partial t}\frac{\partial u_{si}}{\partial t}\right]+
2\rho_{12}^n\left[\frac{\partial u_{i}}{\partial t}\frac{\partial u_{ni}}{\partial t}\right]\right);
\end{eqnarray}
here we take into account that the directions $x, y, z$ are  equivalent and
uncoupled dynamically. We can express $\rho_{11}$, $\rho_{22}^s$, $\rho_{22}^n$, $\rho_{12}^s$ and $ \rho_{12}^n$ in terms of directly measurable coefficients.  Calculations quite similar to [1] for
$\rho_{11}$, $\rho_{22}^s$, $\rho_{22}^n$, $\rho_{12}^s$, $\rho_{12}^n$ density
coefficients give (see also [4]):
\begin{eqnarray}
\rho_{11}&=&(1-\beta)\rho_{sol}-\rho_{12}^s-\rho_{12}^n;\nonumber\\
\rho_{22}^s&=&\beta\rho_{s}-\rho_{12}^s;\nonumber\\
\rho_{22}^n&=&\beta\rho_{n}-\rho_{12}^n;\nonumber\\
\rho_{12}^s&=&(1-\alpha)\beta\rho_{s};\nonumber\\
\rho_{12}^n&=&(1-\alpha)\beta\rho_{n};
\end{eqnarray}
From (23) one can calculate $\rho_{11}$, $\rho_{22}^s$, $\rho_{22}^n$, $\rho_{12}^s$ and $ \rho_{12}^n$ using the  directly measurable coefficients.
$\beta$ is the porosity(fluid volume fraction); $\rho_{sol}$ is the density
of the solid; $\alpha$ is a  purely geometrical quantity independent of solid
and fluid densities; $\alpha=n^2$ where $n$ is the
index of refraction [4].

The $\rho_{12}^s(\rho_{12}^n)$ coefficient represents a mass coupling parameter
between the solid and the superfluid(normal) components.
As $\rho_{s}$ and $\rho_{n}$ can move independently,
the mass 
coupling parameter between the superfluid and the normal 
components is identically
zero.

Let us assume that system is in equilibrium being at rest, therefore any
displacement is a deviation from a state of minimum potential energy. Then
the potential energy $W$ per unit volume of the system is given by [1]:
\begin{equation}
2W=\sigma_xe_x+\sigma_ye_y+\sigma_ze_z+\tau_x\gamma_x+\tau_y\gamma_y+
\tau_z\gamma_z+\sigma_s\epsilon_s+\sigma_n\epsilon_n;
\end{equation}
here, the generalized  procedure of the classical theory of elasticity has been
used [9]. $W$ is expressed
in terms of the strain and stress components.

The stress tensor in  the porous material is:
\begin{displaymath}
\left(\begin{array}{ccc}
\sigma_{x}+\sigma_{s}+\sigma_{n}&\tau_z&\tau_y\\
\tau_z&\sigma_y+\sigma_s+\sigma_n&\tau_x\\
\tau_y&\tau_x&\sigma_z+\sigma_s+\sigma_n
\end{array}\right)
\end{displaymath}
$\sigma_x, \sigma_y, \sigma_z, \tau_x, \tau_y, \tau_z$
components of the stress tensor are the forces applied to the
portion of the cube faces occupied by the solid. Note, that the fraction
of solid area per unit cross section is $(1-\beta)$ [1].
In stress tensor
we formally separate the forces $\sigma_s$ and $\sigma_n$ acting respectively
on the
normal component and the superfluid component "parts" of each face of the cube.
 The "fraction" of the superfluid(normal) component area per unit cross section
is $\beta\rho_s/\rho_h\quad(\beta\rho_n/\rho_h)$. Besides, $\sigma=\sigma_s+\sigma_n$; $\sigma$ is the force
acting on the fluid part of each face of the cube. The fraction of the
fluid area   per unit cross section is $\beta$.

The strain tensor in the solid is:
\begin{displaymath}
\left(\begin{array}{ccc}
e_x&\frac{1}{2}\gamma_z&\frac{1}{2}\gamma_y\\
\frac{1}{2}\gamma_z&e_y&\frac{1}{2}\gamma_z\\
\frac{1}{2}\gamma_y&\frac{1}{2}\gamma_x&e_z
\end{array}\right)
\end{displaymath}
with $e_x=\partial u_x/\partial x$ etc. and $\gamma_z=\partial u_x/\partial y+
\partial u_y/\partial x$ etc.

In the first approximation the stress components are linear functions of the
strain components. The stress-strain relations for superfluid-filled,
statistically isotropic porous media can be expressed as [1]:
\begin{eqnarray}
\sigma_x&=&2Ne_x+Ae+Q_s\epsilon_s+Q_n\epsilon_n;\nonumber\\
\sigma_y&=&2Ne_y+Ae+Q_s\epsilon_s+Q_n\epsilon_n;\nonumber\\
\sigma_z&=&2Ne_z+Ae+Q_s\epsilon_s+Q_n\epsilon_n;\nonumber\\
\tau_x&=&N\gamma_x;\ \tau_y=N\gamma_y;\ \tau_z=N\gamma_z;\nonumber\\
\sigma_s&=&Q_se+R_s\epsilon_s+R_{sn}\epsilon_n;\nonumber\\
\sigma_n&=&Q_ne+R_{sn}\epsilon_s+R_n\epsilon_n;
\end{eqnarray}
where $$e=e_x+e_y+e_z.$$

Note that, because of (25) describes elastic properties of  coupled system of the porous solid and He II, the coefficients  $R_s$, $R_n$, $R_{sn}$ differ from  $R_s^h$, $R_n^h$, $R_{sn}^h$. But the  relations derived in Section II  are still satisfied. Indeed, it is clear that one can consider porous solid as a vessel. So, we can identify  ``unjacketed compressibility test'' and ``second compressibility test''  in Section II  with the unjacketed compressibility test and  the second jacketed compressibility test considered in the  next section respectively. In Section IV we use some of the relations derived in Section II.

It should be pointed out that thermodynamic parameters of He II  in the porous      media   and the  ones of bulk He II may differ from each other                    ($\rho_n$ for example) [3]. 

The thermal expansion coefficient of He II is too  small and we can neglect it. Besides, we neglect the heat exchange between the solid  and He II. This is a  reasonable assumption if the wave     frequency is not too small. Under this  assumptions  it is clear that between the solid and He II forces caused by  temperature gradients do  not exist. So, forces between the solid and He II have ``mechanical   origin''. The two last terms in the first  equation from (25), for instance,  describe the interaction of    the superfluid component and the normal component  with the solid [1]. Because of He II interacts with the porous  solid ``as a whole''  we have:
\begin{displaymath}
Q_s\epsilon_s+Q_n\epsilon_n=Q\epsilon;
\end{displaymath}
and using (4)  one gets:
\begin{equation}
Q_s=\frac{\rho_s}{\rho_h}Q;\quad Q_n=\frac{\rho_n}{\rho_h}Q;
\end{equation}

Now we are ready to write down the general wave  equations.
The calculations following the procedure of ref.[1]
 gives the  equations for
the wave propagation in a superfluid-saturated porous media. Using
(22),(25) and (26)  from Lagrange's equations we derive:
$$
N\vec{\nabla}^2\vec{u}+(A+N)\vec{\nabla}e+\frac{\rho_s}{\rho_h}Q\vec{\nabla}
\epsilon_s+
\frac{\rho_n}{\rho_h}Q\vec{\nabla}\epsilon_n=
$$
$$
=\frac{\partial^2}{\partial t^2}(\rho_{11}\vec{u}+\rho_{12}^s\vec{u}_s+
\rho_{12}^n\vec{u}_n);
$$
\begin{eqnarray}
\frac{\rho_s}{\rho_h}Q\vec{\nabla}e+R_s\vec{\nabla}\epsilon_s+R_{sn}
\vec{\nabla}
\epsilon_n&=&
\frac{\partial^2}{\partial t^2}(\rho_{12}^s\vec{u}+\rho_{22}^s\vec{u}_s);
\nonumber\\
\frac{\rho_n}{\rho_h}Q\vec{\nabla}e+R_n\vec{\nabla}\epsilon_n+
R_{sn}\vec{\nabla}\epsilon_s&=&
\frac{\partial^2}{\partial t^2}(\rho_{12}^n\vec{u}+\rho_{22}^n\vec{u}_n);
\end{eqnarray}
here $\vec{\nabla}$ is the nabla operator.

Because of the statistical isotropy of the material, the rotational waves are
separated from the longitudinal waves and obey independent equations of
propagation [1].

Applying the divergence operator to the set of equations (27) we
obtain:
\begin{eqnarray}
\vec{\nabla}^2\left(Pe+\frac{\rho_s}{\rho_h}Q\epsilon_s+
\frac{\rho_n}{\rho_h}
Q\epsilon_n\right)&=&
\frac{\partial^2}{\partial t^2}(\rho_{11}e+\rho_{12}^s\epsilon_s+
\rho_{12}^n\epsilon_n);
\nonumber\\
\vec{\nabla}^2\left(\frac{\rho_s}{\rho_h}Qe+R_s\epsilon_s+
R_{sn}\epsilon_n\right)&=&
\frac{\partial^2}{\partial t^2}(\rho_{12}^se+\rho_{22}^s\epsilon_s);
\nonumber\\
\vec{\nabla}^2\left(\frac{\rho_n}{\rho_h}Qe+
R_{sn}\epsilon_s+R_n\epsilon_n\right)
&=&\frac{\partial^2}{\partial t^2}(\rho_{12}^ne+\rho_{22}^n
\epsilon_n);
\end{eqnarray}
with $$P=A+2N$$
Equations (28) govern the propagation of longitudinal waves.

Applying the $rot$ operator to equations (27) we obtain:
\begin{eqnarray}
N\vec{\nabla}^2\vec{\omega}&=&
\frac{\partial^2}{\partial t^2}(\rho_{11}\vec{\omega}+
\rho_{12}^s\vec{\omega}_s+\rho_{12}^n\vec{\omega}_n);\nonumber\\
0&=&\frac{\partial^2}{\partial t^2}(\rho_{12}^s\vec{\omega}+
\rho_{22}^s\vec{\omega}_s);\nonumber\\
0&=&\frac{\partial^2}{\partial t^2}(\rho_{12}^n\vec{\omega}
+\rho_{22}^n\vec{\omega}_n);
\end{eqnarray}
with the definitions 
$$\vec{\omega}=rot\ \vec{u};\quad
\vec{\omega}_s=rot\ \vec{u}_s;\quad
\vec{\omega}_n=rot\ \vec{u}_n;$$
Equations (29)  govern the propagation of pure rotational(shear)
waves.

When  $\rho_s=0$ or $\rho_n=0$
equations (28), (29) reduces to the set of
equations obtained by Biot [1]. Besides when $\beta=1$ and $K_B=N=0$ (the case of bulk He II) (28), (29) reduce to (19).

Equations (28),(29) describe the propagation of
longitudinal
and shear waves in a superfluid-saturated, homogeneous and isotropic
porous media in the long wavelength limit when all damping processes
are neglected. 

Note that using results the  of [1] and [5] one can easily incorporate  the  effect of  the wave attenuation due to finite viscousity of the normal component in these equations.

\section{Calculation of the  coefficients }

To make the theory useful we must express coefficients $N,A,Q,R_s,R_n,R_{sn}$ in terms of directly measurable ones [8].

The coefficient $N$ represents the shear modulus of the bulk material,
which is 
directly measurable. $A,Q,R_s,R_n,R_{sn}$ can be expressed in terms
of directly measurable coefficients by considering three compressibility
tests. In the absence of  electrochemical
interfacial effect between fluid and solid $A,Q,R_{sn},
R_s,R_n$  and $N$ are independent
of what fluid is in the pores, including vacuum [4];
we assumed this through
this paper.

In the ``unjacketed'' compressibility test, a sample of the material is immersed in a fluid to which  a pressure $p$ is applied (see Fig. 3).
 Then the adiabatic  bulk modulus of solid $K_S$
is determined by:
\begin{equation}
\frac{1}{K_S}=-\frac{e}{p}.
\end{equation}

In this case (6) takes place, thus fluid pressure has only ``mechanical origin'' and in (25) we have [8]:
\begin{eqnarray}
\sigma_x&=&\sigma_y=\sigma_z=-(1-\beta)p;\nonumber\\
\sigma_s&=&-\beta\frac{\rho_s}{\rho_h}p;\nonumber\\
\sigma_n&=&-\beta\frac{\rho_n}{\rho_h}p.
\end{eqnarray}
And from (25),(26) and (31) we get:
\begin{eqnarray}
-(1-\beta)p&=&\frac{2}{3}Ne+Ae+\frac{\rho_s}{\rho_h}Q\epsilon_s+\frac{\rho_n}{\rho_h}Q\epsilon_n;\nonumber\\
-\beta\frac{\rho_s}{\rho_h}p&=&\frac{\rho_s}{\rho_h}Qe+R_s\epsilon_s+R_{sn}\epsilon_n;  \nonumber\\
-\beta\frac{\rho_n}{\rho_h}p&=&\frac{\rho_n}{\rho_h}Qe+R_{sn}\epsilon_s+R_n\epsilon_n;
\end{eqnarray}

Using (30) from (32) we get:
\begin{equation}
K_S=\frac{\left(A+(2/3)N\right)Z_1-Q^2Z_2}{(1-\beta)Z_1-\beta QZ_2};
\end{equation}
where
\begin{eqnarray}
Z_1&=&R_sR_n-(R_{sn})^2;\nonumber\\
Z_2&=&\frac{\rho_s^2}{\rho_h^2}R_n-2\frac{\rho_s\rho_n}{\rho_h^2}R_{sn}+\frac{\rho_n^2}{\rho_h^2}R_s;
\end{eqnarray}

Besides, (30) and the  two last  equations  from (32) give:
\begin{eqnarray}
R_s&=&\left(\beta-\frac{Q}{K_S}\right)\frac{\rho_s}{\rho_h}K_f-R_{sn};
\nonumber\\
R_n&=&\left(\beta-\frac{Q}{K_S}\right)\frac{\rho_n}{\rho_h}K_f-R_{sn};
\end{eqnarray}
here we have  used (6) and  (9) as well.

In the first ``jacketed'' compressibility test, a sample of the material is
enclosed in a thin impermeable jacket and then subjected to an external
fluid pressure $p$. The fluid inside the jacket can escape
through a tube in the reservoir of He II (see Fig. 4), so the  fluid pressure is exactly zero. The adiabatic  bulk modulus of the skeletal frame $K_B$ is:
\begin{equation}
\frac{1}{K_B}=-\frac{e}{p};
\end{equation}
and in (25) we have [8]:
\begin{eqnarray}
\sigma_x&=&\sigma_y=\sigma_z=-p;\nonumber\\
\sigma_s&=&\sigma_n=0.
\end{eqnarray}
And from (25),(26) and (37) we have:
\begin{eqnarray}
-p&=&\frac{2}{3}Ne+Ae+\frac{\rho_s}{\rho_h}Q\epsilon_s+\frac{\rho_n}{\rho_h}Q\epsilon_n;\nonumber\\
0&=&\frac{\rho_s}{\rho_h}Qe+R_s\epsilon_s+R_{sn}\epsilon_n;  \nonumber\\
0&=&\frac{\rho_n}{\rho_h}Qe+R_{sn}\epsilon_s+R_n\epsilon_n;
\end{eqnarray}

Using (36) from (38) we get:
\begin{equation}
K_B=\frac{(A+(2/3)N)Z_1-Q^2Z_2}{Z_1};
\end{equation}
where $Z_1$ and $Z_2$ are given by (34).

In the second ``jacketed'' compressibility test only the superfluid component is
allowed to escape through a superleak in the reservoir of He II (see Fig. 5). An applied external pressure squeeze out some of the superfluid component; so the temperature of He II in the pores  increases  and mechano-caloric effect [3] takes place.   In this case the bulk modulus of the skeletal frame with the normal fluid component $K_H$ is defined by:
\begin{equation}
\frac{1}{K_H}=-\frac{e}{p}.
\end{equation}
Since only  the superfluid component is allowed to escape through the superleak, the  effective frame here includes both the  solid and the normal fluid component. Because of  the superfluid component and the normal component can not be physically separated, the definition of the bulk modulus of the effective frame  is formal. But let us point out that $K_H$ does not enter into final equations (44); so this problem is irrelevant in this case. 

Let us note  again that we neglect the heat exchange between the porous solid and He II. In general this is not the case. But recall that  we consider the wave processes in a porous medium. And, if the wave frequency is not too small, the heat exchange between the porous solid and He II during times of order of the wave period  is negligible (adiabatic process). Besides, we neglect the thermal expansion coefficient of He II (see Section II).  So, the  response of the sample on the external pressure $p$ simply is a sum of the purely  ``mechanical response'' of the solid and the purely  ``thermal response'' of the normal fluid component. Thus, we can  represent  the external pressure as follows: $p=(p-\beta p')+\beta p'$. The first term in   this sum corresponds to the purely  ``mechanical response''  of the solid, and the second term in this sum  corresponds to the purely  ``thermal response'' of the normal fluid component. In  this case in (25) we have:
\begin{eqnarray}
\sigma_z&=&\sigma_y=\sigma_z=-(1-\beta')p;\nonumber\\
\sigma_s&=&0;\nonumber\\
\sigma_n&=&-\beta'p;
\end{eqnarray}
here $\beta'=\beta(p'/p)$ is the measure of the ``thermal response'' of the      effective frame. If we neglect the mechano-caloric effect $\beta'=0$ and this case reduces to the  first compressibility test.

From (25),(26) and (41) we get:
\begin{eqnarray}
-(1-\beta')p&=&\frac{2}{3}Ne+Ae+\frac{\rho_s}{\rho_h}Q\epsilon_s+\frac{\rho_n}{\rho_h}Q\epsilon_n;\nonumber\\
0&=&\frac{\rho_s}{\rho_h}Qe+R_s\epsilon_s+R_{sn}\epsilon_n;  \nonumber\\
-\beta'p&=&\frac{\rho_n}{\rho_h}Qe+R_{sn}\epsilon_s+R_n\epsilon_n;
\end{eqnarray}

Using (40) and two last  equations from
(42) we can write:
\begin{eqnarray}
0&=&-\frac{\rho_s}{\rho_h}Q\frac{p}{K_H}+R_s\frac{R_{sn}^h}{R_s^h}
\frac{p'}{K_n}-R_{sn}\frac{p'}{K_n};\nonumber\\
-\beta'p&=&-\frac{\rho_n}{\rho_h}Q\frac{p}{K_H}+R_{sn}\frac{R_{sn}^h}{R_s^h}
\frac{p'}{K_n}-R_n\frac{p'}{K_n};
\end{eqnarray}
here we have used (13) and (12) as well.

Finally (33),(39) with (35),(43) and (17),(18) directly  give:
\begin{eqnarray}
A&=&\frac{(1-\beta)\left(1-\beta-(K_B/K_S)\right)+\beta(K_S/K_f)K_B}
{1-\beta -(K_B/K_S)+\beta(K_S/K_f)}-\frac{2}{3}N;\nonumber\\
Q&=&\frac{\left(1-\beta-(K_B/K_S)\right)\beta K_S}
{1-\beta-(K_B/K_S)+\beta(K_S/K_f)};\nonumber\\
R_s&=&\frac{\beta^2K_S}{1-\beta-(K_B/K_S)+\beta(K_S/K_f)}\ \frac{\rho_s^2}{\rho_h^2}\ +\beta
\frac{\rho_hS^2T}{C}\frac{\rho_s^2}{\rho_h^2};\nonumber\\
R_n&=&\frac{\beta^2K_S}{1-\beta-(K_B/K_S)+\beta(K_S/K_f)}\ \frac{\rho_n^2}{\rho_h^2}\ +\beta
\frac{\rho_hS^2T}{C}\frac{\rho_s^2}{\rho_h^2};\nonumber\\
R_{sn}&=&\frac{\beta^2K_S}{1-\beta-(K_B/K_S)+\beta(K_S/K_f)}\frac{\rho_s\rho_n}{\rho_h^2}-\beta
\frac{\rho_hS^2T}{C}\frac{\rho_s^2}{\rho_h^2};
\end{eqnarray}
where $K_f,K_S,K_B$ and $N$ are the directly measurable coefficients. This is the desired result.

When  $\rho_s=0$ or $\rho_n=0$,
(44)  reduces to the set of
equations obtained by Biot and Willis [8]. Besides, when $\beta=1$ and $K_B=N=0$ (the case of bulk He II), equation  (44) reduces to (18).

\section{Particular solutions}

Equations (28) and  (29) describe the propagation of
longitudinal
and shear waves in a superfluid-saturated, homogeneous and isotropic
porous media in the long wavelength limit when all damping processes
are neglected. The general  solutions of these equations will be discussed elsewhere. Here, we consider only the most important particular solutions when 
\begin{equation}
\vec{u}=\vec{u}_n.
\end{equation}
Physically this means that the normal fluid component is locked inside a porous medium by viscous forces and therefore there  is no relative motion
between the normal fluid component and the porous solid. This case is realized in  high-porosity aerogels, for example.

Because of (45) we added together the first and last  equations
of (28). Then, we obtain:
$$
\vec{\nabla}^2\left([P+2\frac{\rho_n}{\rho_h}Q+R_n]e+
[\frac{\rho_s}{\rho_h}Q+R_{sn}]\epsilon_s\right)=
$$
$$
=\frac{\partial^2}{\partial t^2}\left([\rho_{11}+2\rho_{12}^n+\rho_{22}^n]e+
\rho_{12}^s\epsilon_s\right);
$$
\begin{equation}
\vec{\nabla}^2\left([\frac{\rho_s}{\rho_h}Q+R_{sn}]e+R_s\epsilon_s\right)
=\frac{\partial^2}{\partial t^2}\left(\rho_{12}^se+\rho_{22}^s
\epsilon_s\right).
\end{equation}
Similarly, from (29) and (45) we obtain:
\begin{eqnarray}
N\vec{\nabla}^2\vec{\omega}&=&\frac{\partial^2}{\partial t^2}
\left([\rho_{11}+2\rho_{12}^n+\rho_{22}^n]\vec{\omega}+
\rho_{12}^s\vec{\omega}_s\right);\nonumber\\
0&=&\frac{\partial^2}{\partial t^2}(\rho_{12}^s\vec{\omega}+
\rho_{22}^s\vec{\omega}_s).
\end{eqnarray}

Equations (46) and (47) describe
the propagation of longitudinal
and shear waves respectively, when the normal fluid component is
locked inside a porous solid.

It is convenient to introduce the following  definitions:
\begin{eqnarray}
P'&=&P+2\frac{\rho_n}{\rho_h}Q+R_n;\nonumber\\
Q'&=&\frac{\rho_s}{\rho_h}Q+R_{sn};\nonumber\\
\rho'&=&\rho_{11}+2\rho_{12}^n+\rho_{22}^n.
\end{eqnarray}

Let us consider equations (46) for the longitudinal waves.
The solutions of these equations with $\vec{k}$ wave vector are written
in the form
\begin{eqnarray}
e&=&C_1 \exp[i\vec{k}(\vec{r}+\vec{V}t)];\nonumber\\
\epsilon_s&=&C_2 \exp[i\vec{k}(\vec{r}+\vec{V}t)];
\end{eqnarray}
where  $\vec{V}$ is the velocity of these waves; $C_1$ and $C_2$ are real constants. The velocity $\vec{V}$ is determined by
substituting (49) into (46). There are two modes of longitudinal waves:  the fast and the  slow  waves, with $V_f$ and $V_s$ velocities respectively [1]. The velocities of propagation $V_{f,s}$ of fast and slow waves are given by:
\begin{equation}
V_{f,s}^2=
\frac{G\pm\left[G^2-4\left(\rho'\rho_{22}^s-(\rho_{12}^s)^2\right)
(P'R_s-(Q')^2)\right]^
{\frac{1}{2}}}
{2\left(\rho'\rho_{22}^s-(\rho_{12}^s)^2\right)};
\end{equation}
with $$G=P'\rho_{22}^s+R_s\rho'-2Q'\rho_{12}^s;$$
The fast  wave corresponds to the solid with locked normal fluid component and the superfluid component moving in phase, and the slow wave corresponds to solid with locked normal fluid component and the superfluid component moving in opposite
phase. Using the formula (50) we examine the dependence of $V_{f,s}$ upon the temperature. In Fig. 6, Fig. 7 and Fig. 8  fast and slow wave velocities as a function of temperature  are presented for porosities 0.7, 0.98 and 0.99 respectively (see Section VI for discussion). 

From (46) for fast and slow wave amplitudes we have [1]:
\begin{equation}
C_{2\,f,s}=-D_{f,s}C_{1\,f,s};
\end{equation}
where 
$$D_{f,s}=\frac{\,V_{f,s}^2\,\,\rho'\ -\,P'}{V_{f,s}^2\,\rho_{12}^s-Q'}; $$
Using (15) from (49) and (51) we can write down  expression for temperature gradient oscillations (recall that  $\epsilon_n=e$, in this case):
\begin{equation}
\Delta T_{f,s}=(\Delta T_{f,s})_{max}\exp[i\vec k(\vec r+\vec V_{f,s}t)+i\pi];
\end{equation}
where
$$
(\Delta T_{f,s})_{max}=C_{1\,f,s}\,(1+D_{f,s})\,\frac{\rho_s}{\rho_h}\frac{ST}{C};
$$
$(\Delta T_{f,s})_{max}$ is the temperature oscillations amplitude. The sign of $(\Delta T_{f,s})_{max}$  determines the relative phase of $\Delta T_{f,s}$ and $e$ (see equations (49) and (52)). If $(\Delta T_{f,s})_{max}$ is positive, $\Delta T_{f,s}$ and $e$ oscillate with opposite phase; and if $(\Delta T_{f,s})_{max}$ is negative, $\Delta T_{f,s}$ and $e$ oscillate in phase. It must be noted that
 $(\Delta T_{s})_{max}$ is always positive; $(\Delta T_{f})_{max}$ can be positive or negative.

Using the formula (52) we examine the dependence of $(\Delta T_{f,s})_{max}$ upon the temperature. In Fig. 6, Fig. 7 and Fig. 8  temperature oscillations amplitudes for fast and slow waves  as a function of temperature are  presented for porosities 0.7, 0.98 and 0.99 respectively (see Section VI for discussion).  We see that $(\Delta T_{f})_{max}$ is positive in Fig. 6, but  is negative in Fig. 7 and Fig. 8.

It is interesting that, in contrast with bulk He II (see Section II), even for fast wave $\epsilon_s\not=\epsilon_n(=e)$ [1] in general. So, fast wave as well as slow wave is accompanied by temperature oscillations in general. In the case of bulk He II, when  $\beta=1$ and $K_B=N=0$, equation (52) gives that $\Delta T_f\equiv0$; so temperature oscillations do not accompany  first sound,  as we have expected  (see Section II).  

Let us now  consider equations (47) for the shear waves.
 From (47)
it is easy to see that there is only one type of shear wave [1]. The velocity
of propagation of these waves $V_{sh}$ is
\begin{equation}
V_{sh}=\left[\frac{N}{\rho'
\left(1-(\rho_{12}^s)^2/(\rho'\rho_{22}^s)\right)}
\right]^{\frac{1}{2}}.
\end{equation}
Using the formula (53) we examine the dependence of $V_{sh}$ upon the temperature. In Fig. 6, Fig. 7 and Fig. 8  the shear  wave mode velocity as a function of temperature is  presented for porosities 0.7, 0.98 and 0.99 respectively (see Section VI for discussion). 

Note that, as $\rho_{12}^s\not=0$,
the rotation of the solid with locked normal fluid component
$\vec{\omega}$ is coupled to the rotation of the superfluid component
$\vec{\omega}_s$ according to the relation
\begin{equation}
\vec{\omega}_s=-\frac{\rho_{12}^s}{\rho_{22}^s}\vec{\omega};
\end{equation}
and  therefore, $\vec{\omega}_s\not=0$ [1].
Equation (54) shows that the rotation of the solid with locked
normal fluid component and the superfluid component are in the same direction.

Equations (50) and (53) give the velocities of propagation of
 longitudinal and shear  waves in a superfluid-saturated porous media, when 
the normal fluid component is locked inside the porous
solid. If $K_f, K_S, K_B$ and $N$ will be measured, then one can determine 
the velocities of propagation of  fast, slow and shear waves from (50) and (53).
It should be noted that equations (50) and (53)  differ from the equations obtained by Biot [1] and contain significant new physics.  Indeed, Biot's results do not   contain forces caused by temperature gradients. In fact, in the case of superfluid-filled porous media Biot's results are not  valid. Johnson [4] has shown  that in the rigid frame at low temperatures ($T<1.1 K$) fourth sound is exactly Biot's  slow    wave. However, at low temperatures the  hydrodynamical description   breaks  down [3] and comparison to Biot's  model should be made at higher temperatures [5]. So, we conclude that in all physically interesting cases  Biot's equations are not applicable  to superfluid-filled porous media.

Examples of physical systems to which equations (50) and (53) can be applied are 
superleak of porous vycor glass  and high-porosity aerogels
(e.g. silica aerogels with porosities $\beta>0.9$). Superleak of porous vycor
glass corresponds to a superfluid-solid system in the rigid frame limit
($K_f\ll K_S,K_B,N$).
The most interesting physical system,
 high-porosity aerogel, corresponds to a superfluid-solid
system in the weak frame limit($K_B,N\ll K_S,K_f$).

\section{Discussion and conclusions}

Now, let us discuss the results obtained in the previous sections.
In Section III and Section IV  we derived the general equations (28),(29). These  equations    describe the propagation of longitudinal and shear wave in a superfluid-saturated porous media  when all damping processes are neglected. Equations (28) and  (29)  depend explicitly on the  directly measurable  parameters of the porous solid and He II via relations  (23) and (44). Equations (28), (29) and (23), (44)  reduce to the equations  obtained by Biot [1],[8] when $\rho_s=0$ or $\rho_n=0$.  Besides,   when $\beta=1;K_b=N=0$ these equations describe the sound propagation in bulk He II. From what one concludes that  these equations reduce  correctly to Biot's theory [1] and Landau's theory  of sound propagation in bulk He II [3].

In Section V, we considered the case when the normal fluid component is locked inside a porous media  by viscous forces. It is shown that in this case one shear wave and two longitudinal, fast and slow, waves exist. The velocities of these  waves $V_f$, $V_s$, $V_{sh}$  are obtained. 

Since  equations (46) and (47) include pressures caused by thermal gradients, they are more general in comparison with equations derived by Biot. The only case when (46) and (47)  reduce  to the Biot's equations is  $\rho_n=0$ (see also Section V). 

Using (50), (52) and (53) we have plotted the examples of the wave velocities and  temperature oscillations amplitudes for fast and slow waves as a function of temperature. In Fig. 6, Fig. 7 and Fig. 8 the plots are presented for porosities 0.7, 0.98 and 0.99 respectively.

The bulk modulus and the mass  density  of the  solid we choose arbitrarily as  $K_S=4.07\cdot10^{10} N/m^2$ and $\rho_{sol}=4\cdot10^3 kg/m^3$. For the bulk modulus  of the skeletal frame $K_B$  we have used the following relation $K_B=(1-\beta)^{3.5}K_S$ [10]. Poisson's ratio of the skeletal frame is assumed to be $0.2$, thus for the shear modulus of the  solid we have  $N=(3/4)K_B$. When the  porosity is not small $\alpha$ is given by $\alpha=2-\beta$ [4].

In the present paper the  parameters of bulk  He II [11] are used for numerical estimations. But we note that  the  thermodynamic parameters of HeII  in the porous media and the  ones of bulk HeII may differ from each other ($\rho_n$ for example) [3].

Figure 6 shows the  behaviour of the  fast, slow and shear wave velocities as a function of temperature for porosity $\beta=0.7$. Moreover, the temperature  oscillations amplitudes for fast and slow waves $(\Delta T_f)_{max}$ and $(\Delta T_s)_{max}$ are presented as well. In this paper we have assumed that the  temperature oscillations amplitude is about $\sim0.1K$, so for $C_{1f}$ and $C_{1s}$ we choose $C_{1f}=1$ and $C_{1s}=3\cdot10^{-3}$.

The plots presented in Fig. 6 are typical for $\beta<0.9$. For these  porosities  $K_f\ll K_S, K_B, N$; so we have a superfluid-solid system in the rigid frame limit. In this limit the fast wave and shear wave  velocities  decrease when porosity increases. Besides, the slow wave mode is exactly fourth sound. We can show  this  even  analytically.

In the limit of the rigid  frame the expressions (50)
 for the fast and slow wave
are greatly simplified.
 In this case $K_f\ll K_S,K_B,N$ and therefore
\begin{eqnarray}
V_f^2&=&\left[\frac{K_B+(4/3)N}
{(1-\beta)\rho_{sol}+\beta\rho_h-(1/\alpha)\beta\rho_s}\right];\nonumber\\
V_s^2&=&\frac{1}{\alpha}\left(\frac{\rho_s}{\rho_h}V_1^2+
\frac{\rho_n}{\rho_h}V_2^2\right);
\end{eqnarray}
here $V_1$ and $V_2$ are given by (21).

The slow wave corresponds  only to the oscillation of the  superfluid
component (rigid frame) and $V_s$ is the well-known velocity of fourth sound
[3] renormalized by the tortuous, branching path of pore space [4];
 i.e. fourth sound is exactly Biot's slow mode. 

In the limit $T\rightarrow 0$  when there is no the normal fluid component and we have $(\rho_s/\rho_h)=1$, equations (55) are identical to the corresponding expressions obtained in ref.[4]. 

We see from Fig. 6 that in the fast wave mode $(\Delta T_f)_{max}>0$ in the limit of rigid frame. In the  slow wave mode  always $(\Delta T_s)_{max}>0$ (see Fig. 6, Fig. 7 and Fig. 8). In addition let us note that, in contrast with $V_f$ and $V_{sh}$, slow wave velocity  $V_s$ does not depend on $\rho_{sol}$, $K_B$ and $N$.

In Fig. 7 and Fig. 8 we  present  behaviour of the  fast, slow and shear wave velocities as a function of temperature for porosities $\beta=0.98$ and $\beta=0.99$ respectively. Moreover, the temperature  oscillations amplitudes for fast and slow waves $(\Delta T_f)_{max}$ and $(\Delta T_s)_{max}$ are presented as well. For $C_{1f}$ and $C_{1s}$ we choose $C_{1f}=10^{-1}$ and $C_{1s}=3\cdot10^{-1}$.

The plots presented in Fig. 7 and Fig. 8  are typical for $\beta>0.9$. For these  porosities  $K_B, N\ll K_S, K_f$; so we have a  superfluid-solid system in the weak frame limit. In this limit, in contrast with $\beta<0.9$ (rigid frame), the  slow wave velocity  $V_s$ depends on $\rho_{sol}$, $K_B$ and $N$.

Fig. 7 shows that $V_s$ is less compared with fourth sound velocity, but still is  larger than second sound velocity. 

In Fig. 8 we see that the  behaviour of $V_s$ near $T_\lambda$ is almost the same as in Fig. 7. But at low temperatures $V_s$ becomes considerably less than second sound velocity. For porosities $\beta>0.99$ slow wave velocity $V_s$ increases smoothly when  $\beta$ increases; and  when porosity reaches $\beta=1$ the slow wave mode is exactly second sound.

Fast wave velocity $V_f$ decreases when the porosity increases for $\beta<0.95$ (it becomes even less than first sound velocity); and $V_f$ increases when the porosity increases  for $\beta>0.95$. In Fig. 7 and Fig. 8 $V_f$ is less than first sound velocity. For $\beta=1$ fast wave is exactly first sound.  The  shear wave mode velocity $V_{sh}$ decreases when the porosity increases. Note that in Fig. 7 and Fig. 8 $V_{sh}$ is less than $V_s$. For $\beta=1$ the  shear wave velocity is zero.

From Fig. 7 and Fig. 8 we can see that in the fast wave mode  $(\Delta T_f)_{max}<0$ in the limit of weak frame.  When porosity is exactly $\beta=1$ (the case of  bulk He II), the temperature oscillations amplitude in the  fast wave mode  is zero (also see Section V).

The limit of weak frame is the  most interesting and important application  of our   theory. Indeed, in recent years the investigation of the  behaviour of superfluids in aerogels has attached increasing attention (for example, see [12],[13],[14]). Aerogels are highly open porous solids with  porosities $\beta>0.9$ (e.g. silica aerogels). The mean pore diameter in aerogels is about $20 nm$, so the normal fluid component is clamped in the frame by viscous forces. Besides,  high-porosity aerogels have very small bulk and shear modulus  and are highly   compliant. That is why in the superfluid-filled aerogels  pressures caused by thermal gradients play a  crucial role, in contrast to  other porous solids. Thus, one  should  describe  wave  processes in high-porosity aerogels  using equations (46) and (47).

We  point out that the theory of sound propagation in aerogels has been developed in ref.[12]. These authors modified the conventional two fluid hydrodynamic  equations and derived an expression (see equation (8) in ref.[12])  which gives the longitudinal wave velocity in superfluid-filled  aerogels, but  does not give  the shear wave velocity. It must be noted that  equation (8) in ref.[12] is  incorrect. To show this we remark the  following.  

The longitudinal wave velocity in any solid (in aerogel in the case of ref.[12]) is given by  $[(K_a+(4/3)N_a)\,/\,\rho_a]^{1/2};$ [15]. Here $K_a$, $N_a$  and $\rho_a$ are the bulk modulus, the shear modulus and the  mass density of the solid, respectively. In [12] the  term $(4/3)N_a/\rho_a$ is omitted (see the equation (5) in ref.[12]). This is justified if $N_a/K_a\ll 1$. In the case of aerogels Poisson's ratio equals $\sim0.2$ and is independent of density [16],[10]. This means that $N_a/K_a\sim 1$ and the term $(4/3)N_a/\rho_a$ in  the longitudinal wave velocity is not negligible.

Besides, in the limit $\rho_a\rightarrow\infty$ equation (8) in ref.[12]  suggests  that in the superfluid-filled  porous media one of the modes of sound   propagates  with  velocity  $C_a$, where $C_a$ is given by equation  (5) in ref.[12], which is incorrect (see [15]). 

In conclusion, we have derived the general equations which describe the propagation of longitudinal and shear waves in a superfluid-filled, homogeneous and isotropic porous media in the long wavelength limit when all damping processes are neglected. The most important case when  the normal fluid component is locked  inside a porous media by viscous forces is investigated in detail. It is shown that, in this case, one type of  shear wave and two types of  longitudinal, fast and slow wave, propagation are possible. The velocities of these  waves are obtained. Our results reduce to the  well-known equations in the appropriate limits. We have also demonstrated that the  formalism developed by Biot may be applied to the wave processes in He II. It is shown that fourth sound, as well as second sound in bulk He II is exactly Biot's slow wave. In addition,    we remark that the  results derived in this paper may be applied to   the porous media saturated with superfluid $^3$He as well.

\newpage

FIGURE CAPTIONS :

\vspace{2cm}

Fig. 1. The ``unjacketed compressibility test'' for bulk He II. A pressure $p$ is applied to a volume of  He II in a vessel. In this case none of the fluid components  is allowed to escape from vessel.

\vspace{1cm}

Fig. 2. The ``second jacketed  compressibility test'' for bulk He II. A pressure $p'$ is applied to He II in a vessel. The applied external pressure squeeze out some of the superfluid component through the superleak into the reservoir of He II. As a consequence temperature difference between the fluid in  the vessel and the reservoir of He II  exists.

\vspace{1cm}

Fig. 3. The ``unjacketed'' compressibility test. A pressure $p$ is applied to a sample of the superfluid-filled porous solid. In this case none of the fluid components  is allowed to escape into the reservoir of He II.

\vspace{1cm}

Fig. 4. The first ``jacketed'' compressibility test. A pressure $p$ is applied to a sample of the superfluid-filled porous solid. The applied external pressure squeeze out some of the fluid  through the tube into the reservoir of He II. In this case both of the fluid components are  allowed to escape into the reservoir of He II. Thus, no temperature difference between the superfluid in the porous solid and the reservoir of He II exists.

\vspace{1cm}

Fig. 5. The second ``jacketed''  compressibility test. A pressure $p$ is applied to a sample of the superfluid-filled porous solid. The applied external pressure squeeze out some of the superfluid component through the superleak into the reservoir of He II. In this case only the superfluid component is allowed to escape into the reservoir of He II.  As a consequence temperature difference between the superfluid in the porous solid and the reservoir of He II exists.

\vspace{1cm}

Fig. 6.  The velocities and the temperature oscillations amplitudes as a function of temperature for fast and slow waves; and the shear wave velocity as a function of temperature. $\beta=0.7$, $C_{1f}=1$ and $C_{1s}=0.003$; $K_S=4.07\cdot10^{10} N/m^2$ and $\rho_{sol}=4\cdot10^3 kg/m^3$; $K_B=(1-\beta)^{3.5}K_S$ and $N=(3/4)K_B$;  $\alpha=2-\beta$. Here the  parameters of bulk  He II are used for numerical estimations.

\vspace{1cm}

Fig. 7. The velocities and the temperature oscillations amplitudes as a function of temperature for fast and slow waves; and the shear wave velocity as a function of temperature. $\beta=0.98$, $C_{1f}=0.1$ and $C_{1s}=0.3$; $K_S=4.07\cdot10^{10} N/m^2$ and $\rho_{sol}=4\cdot10^3 kg/m^3$; $K_B=(1-\beta)^{3.5}K_S$ and $N=(3/4)K_B$; $\alpha=2-\beta$. Here the  parameters of bulk  He II  are used for numerical estimations.

\vspace{1cm}

Fig. 8. The velocities and the temperature oscillations amplitudes as a function of temperature for fast and slow waves; and the shear wave velocity as a function of temperature. $\beta=0.99$, $C_{1f}=0.1$ and $C_{1s}=0.3$; $K_S=4.07\cdot10^{10} N/m^2$ and $\rho_{sol}=4\cdot10^3 kg/m^3$; $K_B=(1-\beta)^{3.5}K_S$ and $N=(3/4)K_B$; $\alpha=2-\beta$. Here the  parameters of bulk  He II  are used for numerical estimations.

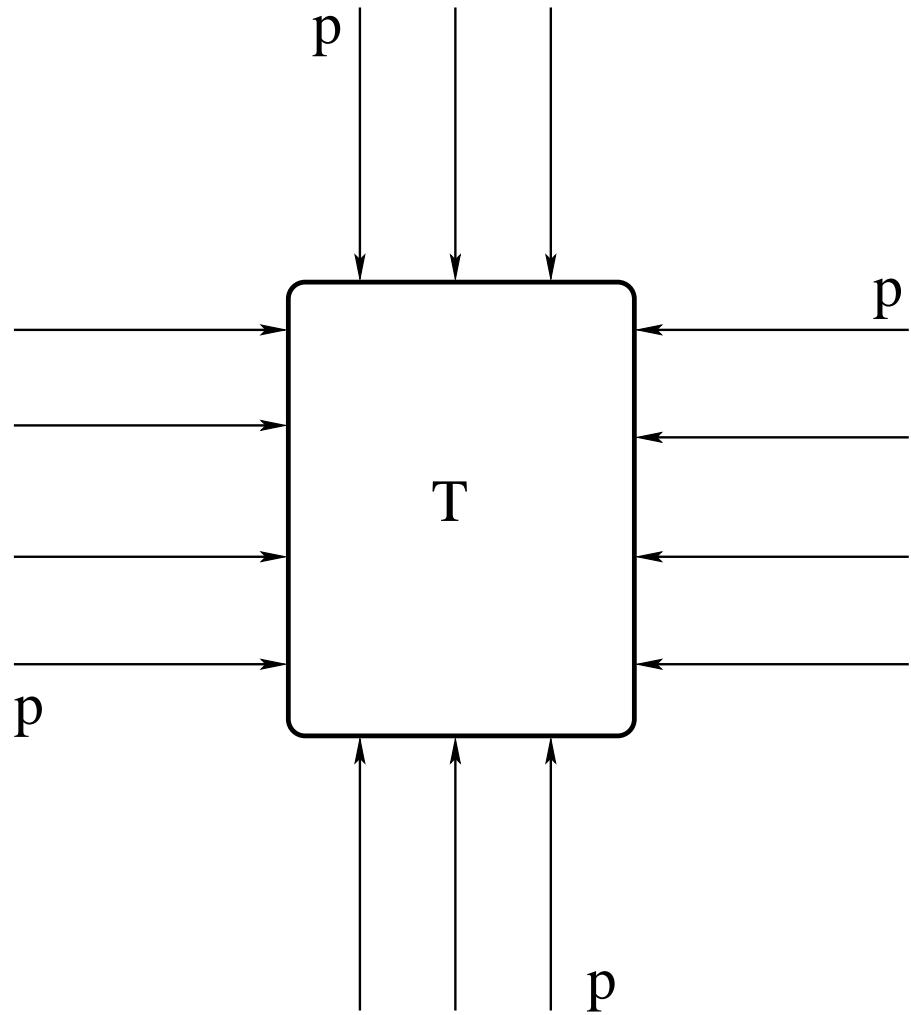
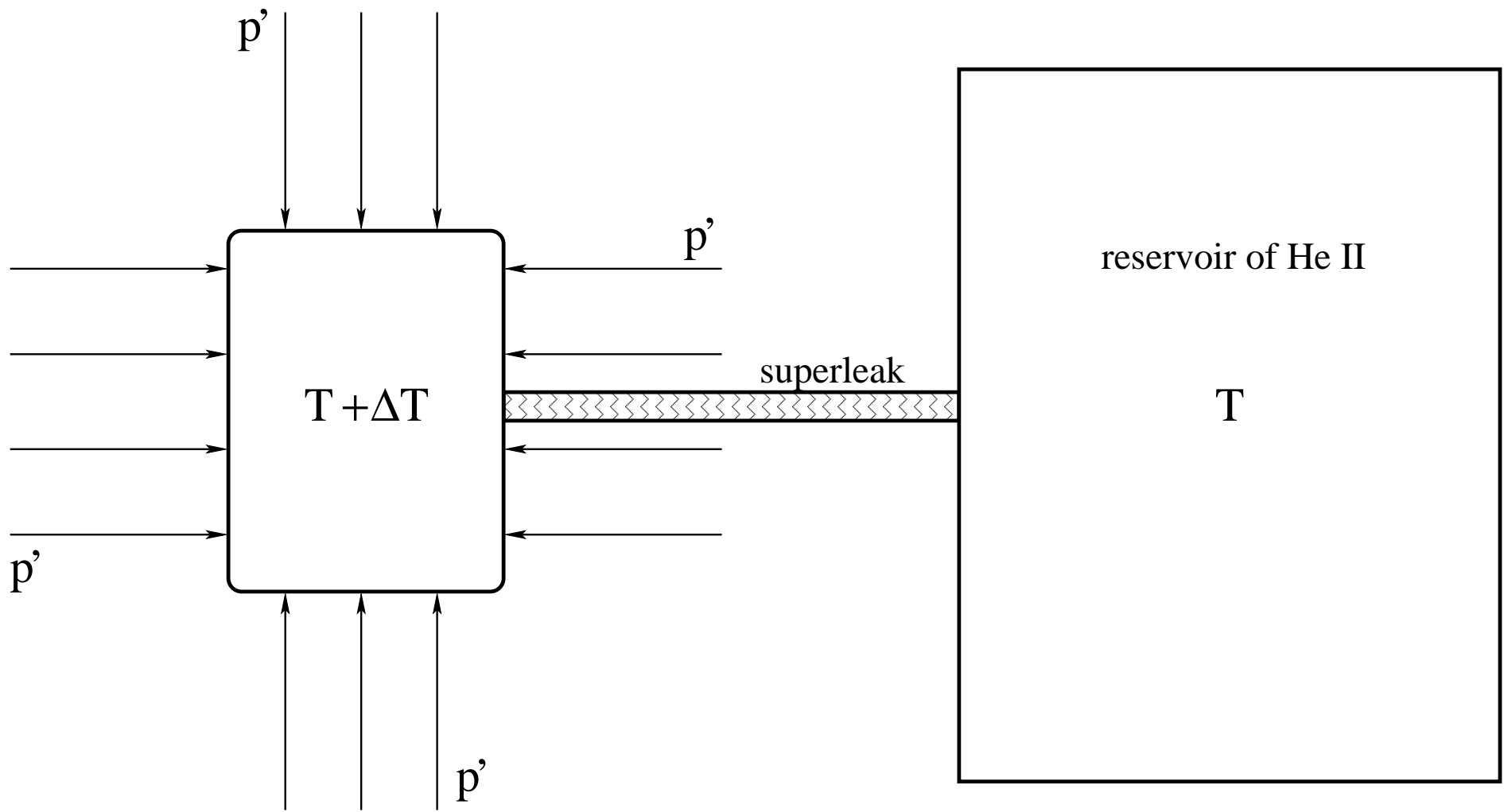
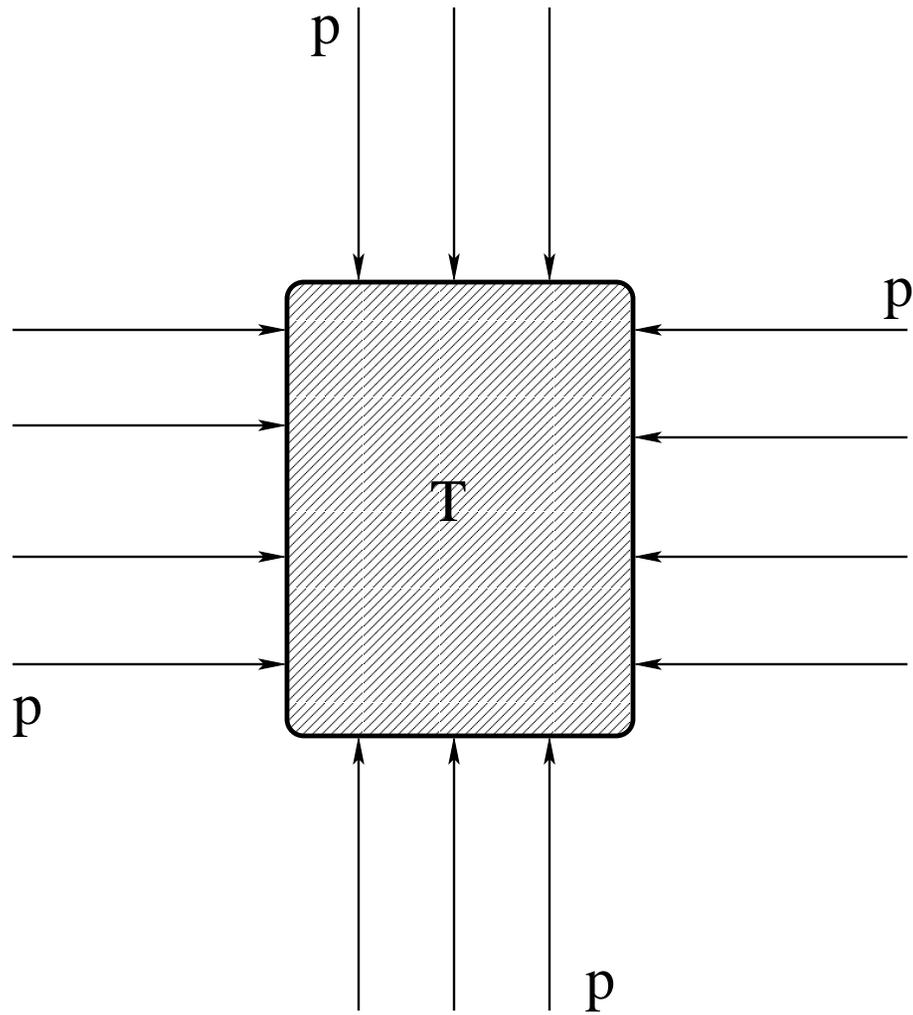
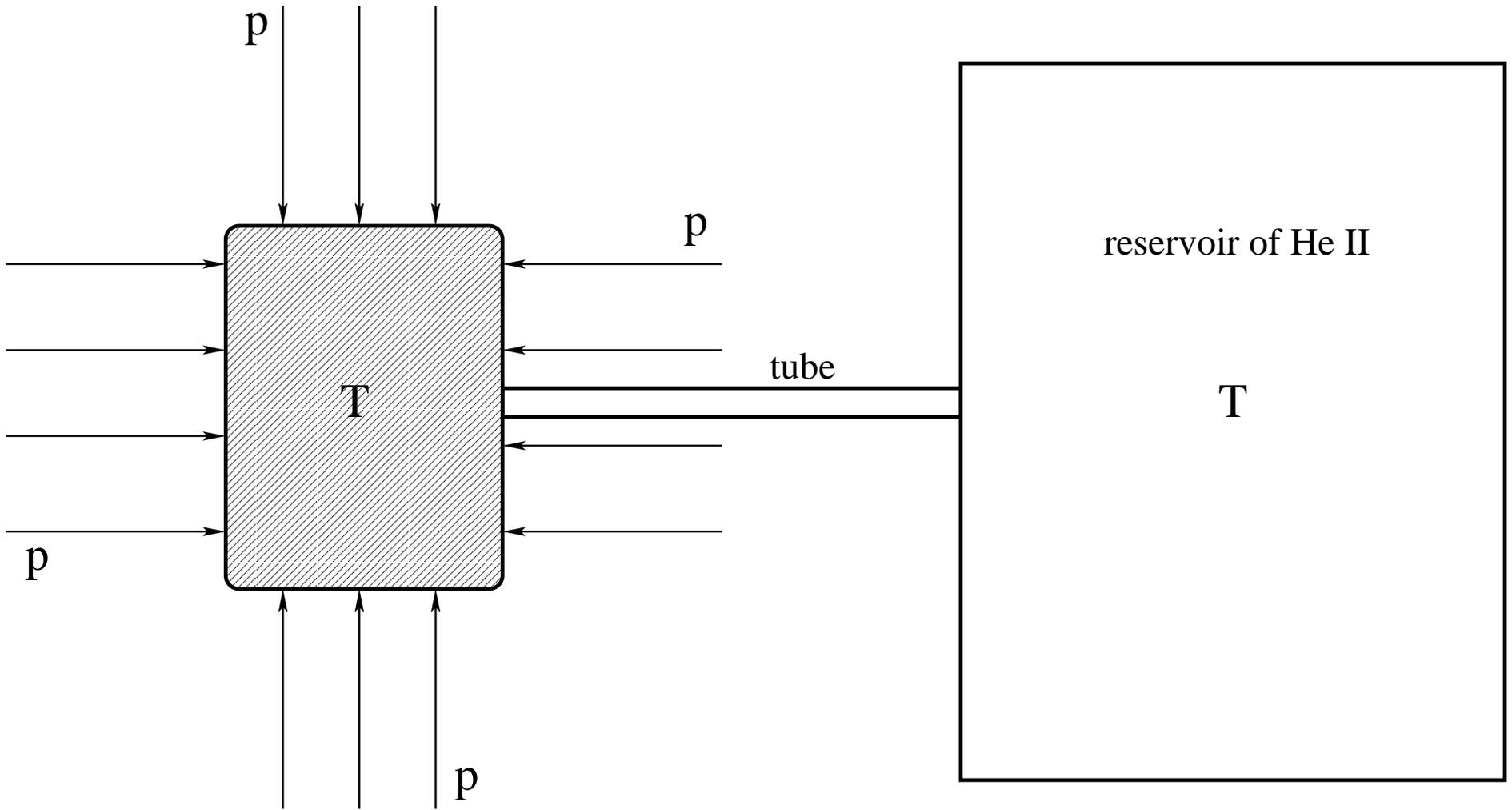
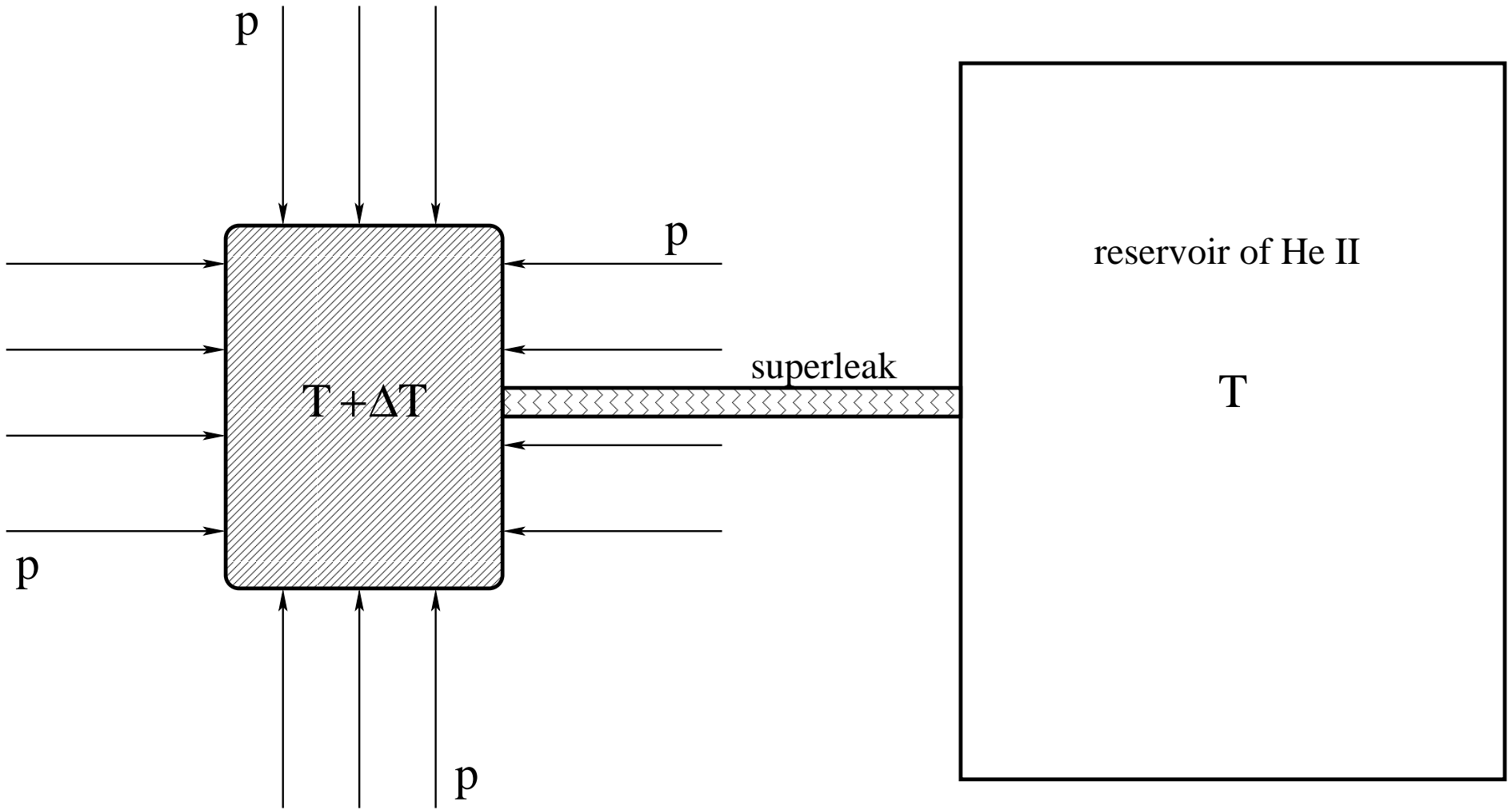
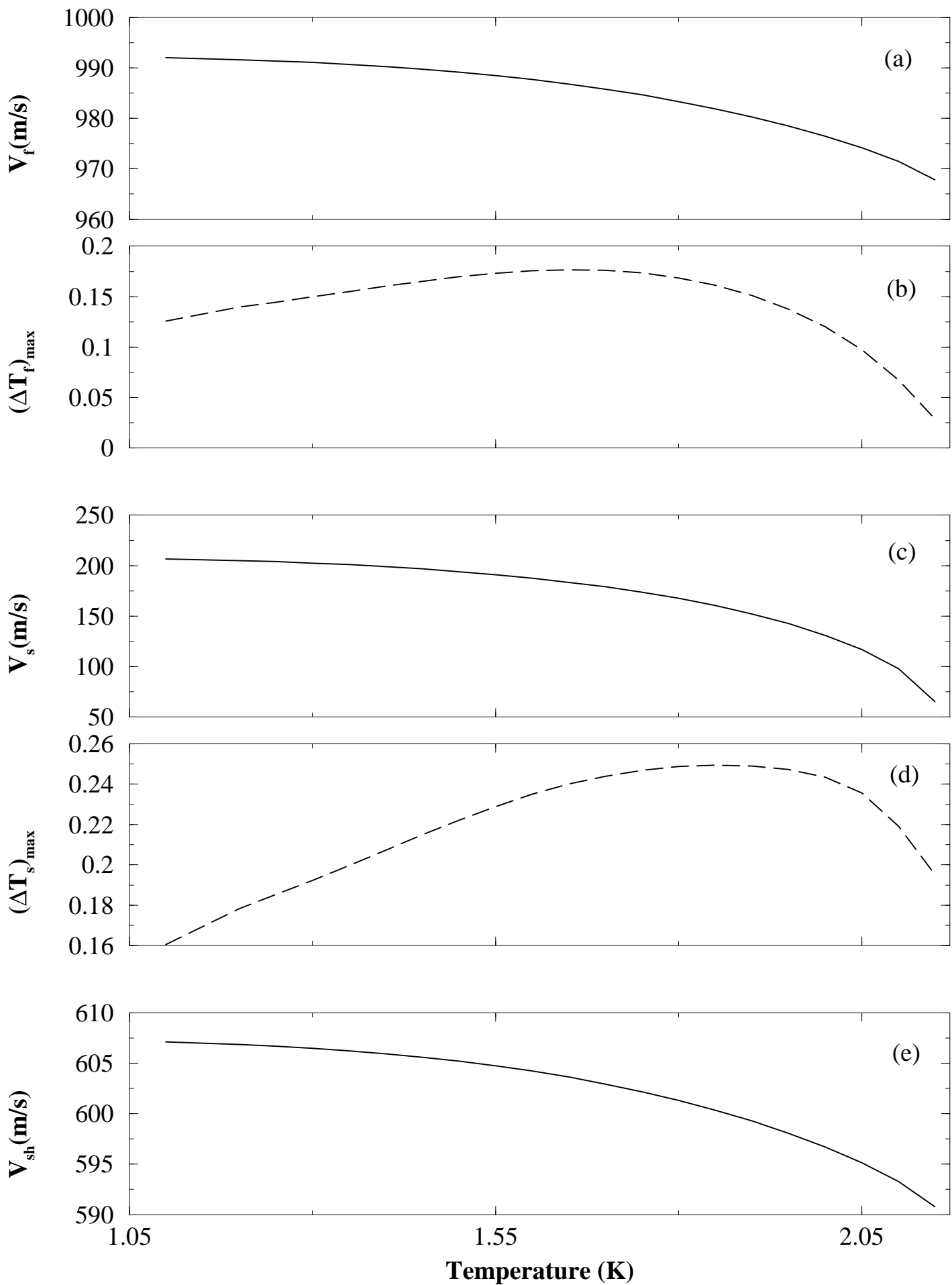
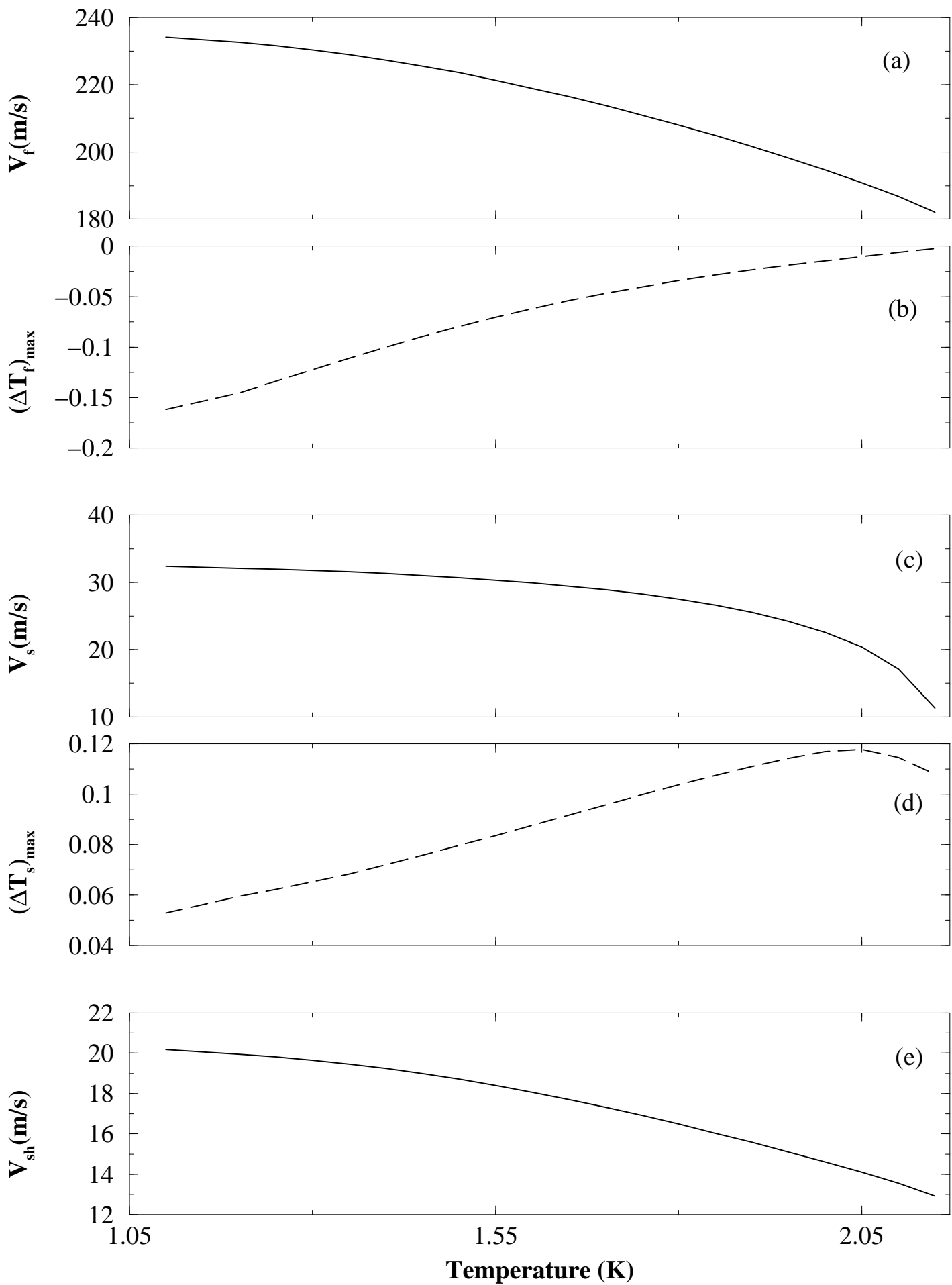
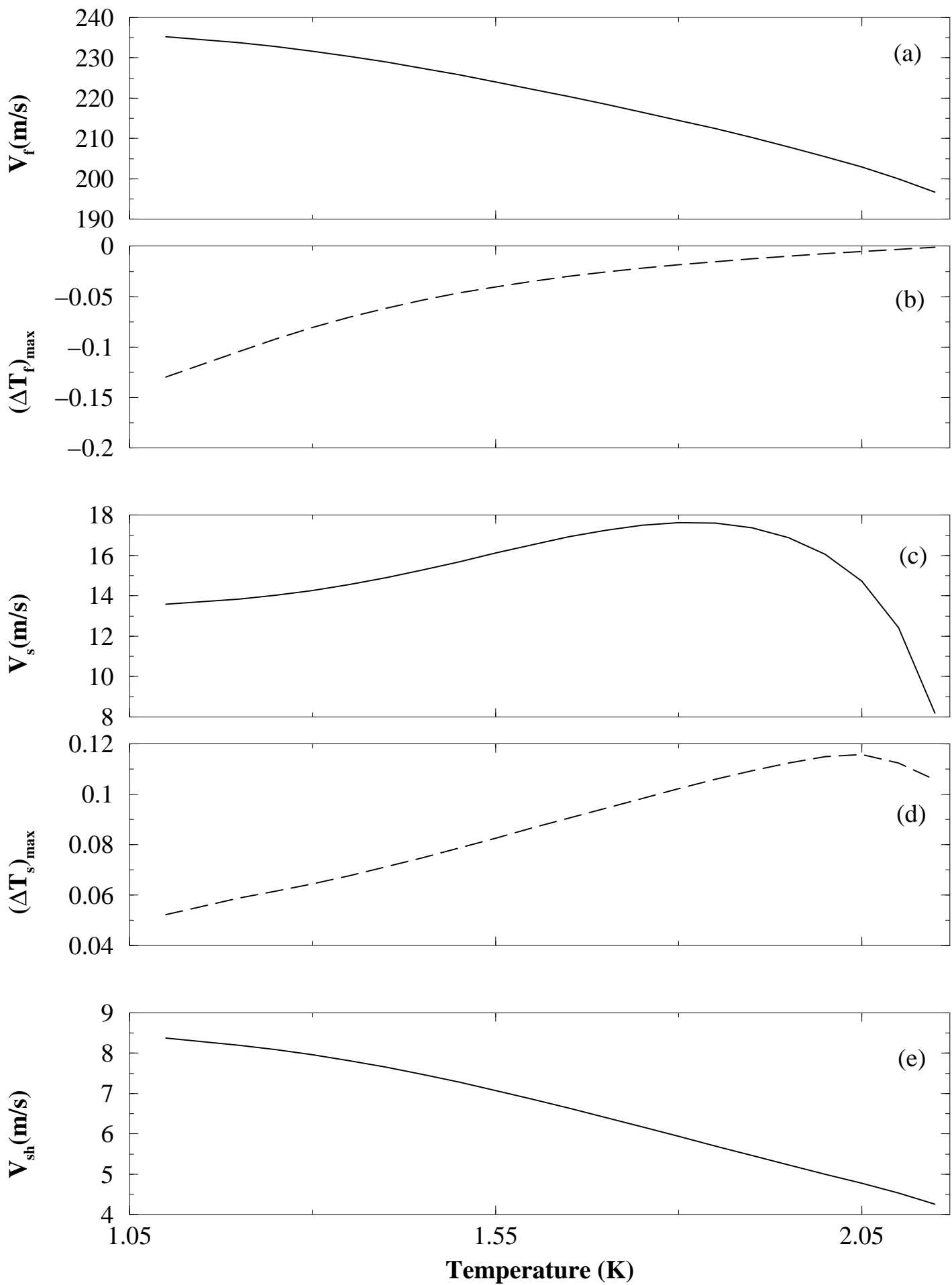
\end{document}